\documentclass{aa}
\usepackage{epsfig}
\usepackage{natbib}
\usepackage{amssymb}
\bibpunct{(}{)}{;}{a}{}{,}

\newcommand{\bez}{\begin{eqnarray*}}
\newcommand{\eez}{\end{eqnarray*}}
\newcommand{\be}{\begin{equation}}
\newcommand{\ee}{\end{equation}}
\newcommand{\beq}{\begin{eqnarray}}
\newcommand{\eeq}{\end{eqnarray}}
\newcommand{\bc}{\begin{center}}
\newcommand{\ec}{\end{center}}

\def \Delnud{\Delta \nu_{\rm D}}

\def \d{{\rm d}}
\def \pc{p^{\rm c}}
\def \fw{f_{\rm H_2O}}
\def \fwd{f_{\rm H_2O}/f_{\rm d}}
\def \fd{f_{\rm d}}
\def \alw{\alpha_{\rm L}}
\def \water{H$_2$O}
\def \ald{\alpha_{\rm d}}

\def \Nh{N_{\rm H_2}}
\def \Nw{N_{\rm H_2O}}
\def\delnm{\Delta n_{\rm m}}
\def\eem{E_{\rm m}}

\def \rhod{\rho_{\rm d}}
\def \Bd{B^{\rm d}}
\def \Bg{B}

\def \Td{T_{\rm d}}

\def \mum{\mu{\rm m}}

\def \taud{\tau_{\rm d}}
\def \tauc{\tau_{\rm c}}
\def \taum{\tau_{\rm m}}
\def \ginv{{\rm g}^{-1}}
\def \gram{{\rm g}}
\def \cminvone{{\rm cm}^{-1}}
\def \cminv3{{\rm cm}^{-3}}

\def \cmtwo{{\rm cm}^{2}}

  \def \TH{T_{\rm H}}
  \def \Te{T_{\rm e}}
\def\ij{{ul}} 
\def\ji{{lu}} 
\def\ul{{ul}} 
\def\lu{{lu}}
\def\up{{u}} 
\def\low{{l}}

\def\Jij{\overline{J_{\ij}}}

\def\J{\overline{J}}

\def\Sul{S_{\ul}}
\def\Cij{C_{\ij}}
\def\Cji{C_{\ji}} 
\def\kij{k_{\ij}} 
\def\Rij{R_{\ij}} 
\def\Rji{R_{\ji}}

\def\Wij{W_{\ij}}
\def\Wji{W_{\ji}}
\def\Rul{R_{\ul}} 
\def\Rlu{R_{\lu}} 
\def\Aij{A_{\ij}}
\def\Aji{A_{\ji}}
\def\Bij{B_{\ij}} 
\def\Sij{S_{\ij}}
\def\pji{p_{\ji}}
\def\pij{p_{\ij}}
\def\pdij{p^{\rm c}_{\ij}}
\def\pdji{p^{\rm c}_{\ji}}

\def\lambdaij{\lambda_{\ij}} 
\def\nuij{\nu_{\ij}} 
\def\nij{n_{\ij}}
\def\nji{n_{\ji}}
\def\ni{n_{\up}}
\def\nj{n_{\low}}
\def\gi{g_{\up}}
\def\gj{g_{\low}}

\def\aul{A_{\ul}}

\def\Rul{R_{\ul}}
\def\Rlu{R_{\lu}}
\def\Cul{C_{\ul}}
\def\Clu{C_{\lu}}

\def\gu{g_{\up}} 
\def\gl{g_{\low}} 
\def\nup{n_{\up}} 
\def\nlow{n_{\low}} 

\def\mp{m_{\rm p}}
\def\kabs{\kappa_{\rm d}}

\begin{document}

\title{Water masers in dusty environments}

\author{Natalia Babkovskaia \and Juri Poutanen}
\institute{Astronomy Division, P.O.Box 3000, FIN-90014 University of Oulu, Finland \\
 \email{natalia.babkovskaia@oulu.fi; juri.poutanen@oulu.fi}}

\titlerunning{Water masers  in dusty environments}
\authorrunning{Natalia Babkovskaia \and Juri Poutanen }


\date{\today}

\abstract{We study in details a pumping mechanism  for the $\lambda=1.35$ cm maser
transition  $6_{16} \rightarrow 5_{23}$ in ortho-\water\ based on the
difference between gas and dust temperatures.
The upper maser level is populated radiatively through
$4_{14} \rightarrow 5_{05}$ and $5_{05} \rightarrow 6_{16}$ transitions.
The heat sink is realized by absorbing the 45 $\mum$ photons, corresponding
to the $5_{23} \rightarrow 4_{14}$ transition, by cold dust.
We  compute the inversion of maser level populations in the optically thick
medium as a function of the hydrogen concentration, the gas-to-dust mass ratio, and
the difference between the gas and the dust temperatures.
The main results of numerical simulations are interpreted
in terms of a simplified four-level model.
We show that the maser strength depends mostly on the product of
hydrogen concentration and the dust-to-water mass ratio  but not on
the size distribution of the dust particles or their type.
We also suggest approximate formulae that describe accurately the inversion 
and can be used for fast calculations of the maser luminosity.
Depending on the gas temperature,
the maximum maser luminosity is reached when the water concentration
$\Nw\approx 10^{6}\div 10^{7}\cminv3$ times the dust-to-hydrogen mass ratio,
and the inversion completely disappears at density just an order of magnitude
larger.  For the dust temperature of 130 K, the $6_{16} \rightarrow 5_{23}$ 
transition becomes
inverted already at the temperature difference of $\Delta T\sim 1$ K,
while other possible masing transitions require a larger $\Delta T\gtrsim 30$ K.
We identify the region of the parameter space where other ortho- and para-water 
masing transitions can appear.
  }

\maketitle


\section{Introduction}

Sources of strong water maser emission at wavelength $\lambda = 1.35$ cm have been
found in many astrophysical objects such as
active galactic nuclei, carbon rich stars,  protostellar regions and comets.
Maser emission is a powerful tool for investigating
the physical conditions in the emitting regions
because of its high brightness as well as a  high sensitivity to
the physical parameters of the medium in which the maser amplification takes place.

\begin{figure}
\centerline{\epsfig{file=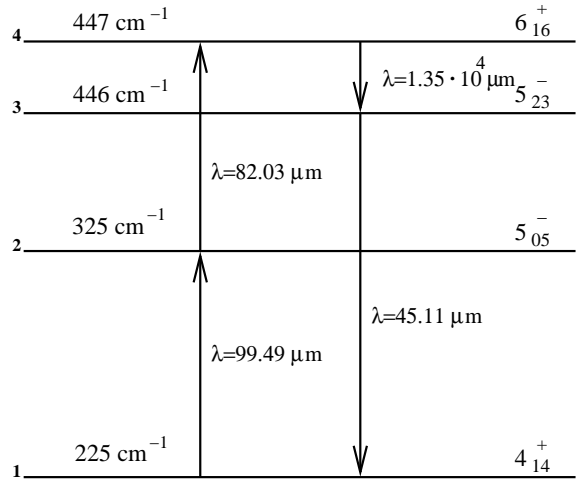,width=7.5cm} }
\caption{A portion of the diagram of the ortho-\water\
showing rotational levels involved in the pumping of $6_{16} \rightarrow 5_{23}$
maser.
}
\label{fig_pump}
\end{figure}

There are several mechanisms for the ortho-\water\ $6_{16} \rightarrow 5_{23}$ 
($\lambda = 1.35$ cm) maser pumping.
\cite{dJ73} proposed a comprehensive model where the breakdown of thermal equilibrium
occurs in the surface layers of the dense gas cloud.
Upon approach to the surface,  the $5_{23}\rightarrow 4_{14}$ transition (45 $\mum$)
(which is among the most important for the maser action, see Fig.~ \ref{fig_pump})
becomes transparent first and level $5_{23}$  becomes underpopulated.
The  de Jong model was widely used to describe the water masers
from late-type stars and star-forming regions \citep{CE85, EH89, NM91}, and
the molecular accretion disks in active galactic nuclei \citep{NMC94,BV00}.

The de Jong mechanism is, however, not able to explain the most powerful
masers in star-forming regions. A different physics has to be involved.
In shocks, the departures from the equilibrium could be produced
by  collisions with hotter ``superthermal'' hydrogen gas \citep{VK83}.
\citet{S80,S84} proposed that collisions with two species
(e.g. charged and neutral particles) of different temperatures
and with comparable collision rate
can produce the necessary inversion at high hydrogen densities
needed to explain the high observed maser luminosities.
\cite{KN87} pointed out that
the magnetohydrodynamic shocks produce naturally conditions where
the  electron temperature $\Te$ is larger than the hydrogen
temperature $\TH$.
In such a case, the lower maser level becomes  underpopulated,
because the relative importance of collisions with neutrals  is larger
for the transitions to/from this level. 
The inversion in this model appeared only in a narrow range
of the ionization fraction $10^{-5}-10^{-4}$ and for $\TH<60$ K.
At high ionization, electrons
start to dominate the collisions and the levels are thermalized at $\Te$, while
at lower ionization, neutrals are dominating and the thermalization happens at $\TH$.
Using better estimates for the collision rates, \citet{EF89} and \citet{AW90}
ruled out this model unless the neutral particles are hotter than the charged
ones, conditions   that are difficult to imagine in any astrophysical
environment.

All astrophysical objects that show water maser emission are also
expected to have non-negligible quantities of dust.
If the dust and the gas have different temperatures,
the departures from the equilibrium are possible \citep{GK74,K75,S77,B77}.
\cite{GK74} suggested that  the radiation from the hot dust
excites the water molecules to the vibrational state, and the heat sink
is provided by collisions with cooler (than dust) hydrogen molecules.
The possibility of the inversion in such a situation
was questioned by \cite{De81} who pointed out
that  because the collisional de-excitation rate between
vibrational states is much smaller than the pure-rotational
collisional rate, the rotational collisional thermalization
may quench the maser when vibrational collisional de-excitation becomes dominant
\citep[see also][]{S88}.

Alternatively, the cold dust can produce the necessary inversion \citep{S77,B77}.
 In the optically thick environment, the excitation
temperature takes the values between the dust and the gas temperatures depending on
the relative role of dust and collisions in 
the destruction of the line photons. 
\cite{De81} considered the following cycle of the maser levels pumping
 $4_{14} \rightarrow 5_{05} \rightarrow 6_{16} \rightarrow 5_{23} \rightarrow 4_{14}$
(see Fig.~\ref{fig_pump}). He showed that 
the downward transition $5_{23} \rightarrow 4_{14}$ at 45 $\mum$ is much more
affected by the dust absorption
than the upward transitions at $\lambda\sim 80-100\ \mum$, because 
there is a  strong peak near 45 $\mum$ in the absorption coefficient 
of the cosmic-type ices \citep[e.g.][]{MH94}.
The upper level excitation temperature is then close
to the gas temperature, while the lower level becomes populated at the
dust temperature.
Similar inversion occurs for other types of dust too.

One should note that in this model an arbitrary
thick layer can participate in the maser action provided the gas and dust
temperatures sufficiently differ. \citet{CK84a} computed the
maser efficiency when both the surface escape and the cold dust absorption mechanisms
operate together.
Recently \citet{CW95} and \citet{WW97} applied this model
to the masing disk in the active galaxy NGC 4258 concluding that
the cold dust model is much more efficient \citep[see also][]{N00}.
Thus the cold dust--hot gas model seems to be the most
promising to explain powerful water masers in many astrophysical objects.
The temperature difference between the gas and the dust
can appear as a result of shock heating or 
illumination by the UV- or X-ray photons,
and/or because of the presence of the dust particles of different types and sizes 
which assure different temperatures.

The purpose of the present study is to determine the
maser strength for a broad range of the physical parameters:
the hydrogen concentration, the water-to-dust mass ratio,
the gas and  the dust temperatures as well as the dust type.
Previous studies \citep{De81,CK84a} used the
rectangular line profile instead of the Doppler one as well as 
the collisional  coefficients with hydrogen of \cite{Gr80}, which
later have been much improved \citep{Gr93}. 
Only a handful set of parameters was explored in 
the more recent investigation of \citet{CW95} and \citet{WW97} 
who considered the case of saturated maser. 

In its full statement this problem needs the simultaneous
solution of   the statistical balance equations together with the
radiative transfer equations for all spectral lines.
However, because the cold dust -- hot gas model can operate inside a
molecular cloud where almost all transitions (except masing)
are optically thick, the radiative transfer can be handled in a very
simple manner using the escape probability formalism, where now the role
of the escape of spectral line photons  is played by the dust absorption.

The paper is constructed as follows.
We give the formulation of  the problem and
describe the numerical method in Sect.~\ref{sec:method}.
We develop a simple four-level model
that contains most of the physics involved in Sect.~\ref{sec:anal}.
The numerical results for different dust types and size distributions
are presented in  Sect.~\ref{sec:result}, where
we also propose simple analytical formulae for the inversion
which describe the results of simulations.
Finally, in  Sect.~\ref{sec:concl} we present our conclusions.

\section{Method}
\label{sec:method}
\subsection{Radiative transfer}

We consider a slab consisting of a mixture of molecular hydrogen,
water molecules and dust particles.
The radiative transfer equation for the specific intensity $I_{x}$
describing the transfer of the line radiation
in the presence of the continuum absorption and emission is \citep{Hu68, Mi78}
\begin{equation}     \label{eq:rte}
\mu \frac{\d I_{x}}{\d z}=-
(\alw \phi_{x} +\ald)I_x+\alw \phi_{x} \Sij+\ald \Bij(\Td),
\end{equation}
where $z$ is the geometrical depth, $\mu$ is the cosine of the 
angle between the direction of propagation and the outward normal,
\be \label{eq:sour}
\Sij=
\left( \frac{\nj}{\ni}\frac{\gi}{\gj}-1 \right)^{-1},
\;\; \up>\low,
\ee
is the source function  for the line transition between the
energy levels $\up$ and $\low$,
and $\Bij(\Td)$ is the Planck function at the transition frequency
$\nuij$ characterized by the dust temperature.
All the intensities and the source functions are given in units 
$2h\nuij  ^3/c^2$.
The dust absorption coefficient  $\ald=\kabs \rhod$
depends on the frequency-dependent (constant within the line)
opacity $\kabs\ [\cmtwo\ginv]$ and
the dust density $\rhod=\fd 2\mp \Nh\ [\gram\ \cminv3]$, where
$\mp$ is the proton mass, $\Nh$  is the concentration of molecular hydrogen, and
$\fd$ is the dust-to-hydrogen mass ratio.

The line averaged absorption coefficient $\alw$ (including induced emission)
depends on the corresponding level populations:
\be   \label{eq:alw}
\alw=\gi \Aij \Delta \nij
\frac{\lambdaij^3}{8\pi c (\Delnud/\nuij) } \Nw ,
\ee
where $\Delta \nij=\frac{\nj}{\gj}- \frac{\ni}{\gi} $,
$\ni$ is the fractional level population ($\sum_\up \ni=1$),
$\Nw$ is the water concentration,
$\gi$ is the statistical weight for a given level,
$\Aij$ is the Einstein coefficient, and $\lambdaij=c/\nuij$ is the transition
wavelength.
We further assume complete frequency redistribution within the line,
i.e. both the absorption and emission line profile are described by the same
function $\phi_x$ normalized as $\int \phi_x \d x=1$, where $x=(\nu-\nuij)/\Delnud$
is the frequency within the spectral line
in thermal Doppler units $\Delnud=\nuij(2kT/mc^2)^{1/2}$.
For the Doppler profile assumed here, $\phi_{x}=\pi^{-1/2}\exp(-x^2)$.

Let us note the averaged over line profile optical thickness of the slab
in $\ij$--line as $2\tau_0=\int \alw \d z$.
A formal solution of the radiative transfer equation (\ref{eq:rte})
for a homogeneous (i.e. $T=const, \Td=const$) 
slab gives  the mean intensity of the radiation averaged
over the line profile at the optical depth $\tau$  \citep{Hu68,Mi78}:
\begin{eqnarray}  \label{SRT}
\Jij (\tau)&=&
\int \phi_x J_x(\tau) \d x=
(1-\delta)\int_0^{2 \tau_0}  \Sij(t) K_1(t - \tau) \d t
\nonumber \\
& + & \delta \Bij(\Td) \int_0^{2 \tau_0} L_1(t - \tau) \d t,
\end{eqnarray}
where the kernels
\begin{eqnarray} \label{eq:KL1}
K_1(s)&=& \frac{1}{2(1-\delta)}
\int E_1\left( [\phi_x+\beta]|s|\right) \phi_x^2 \d x, \\
L_1(s)&=&  \frac{\beta}{2\delta}
\int E_1\left( [\phi_x+\beta]|s|\right) \phi_x \d x
\end{eqnarray}
are  normalized as $\int_{-\infty}^{\infty} K_1(t) \d t=
\int_{-\infty}^{\infty} L_1(t) \d t=1$.
Here $E_n(\tau)=\int_{0}^{1}\mu^{n-2} \exp(-\tau/\mu) \d \mu$ is
the exponential integral function,
\be  \label{eq:beta}
\beta=\ald/\alw,
\ee
and
\be \label{eq:delta}
\delta=\beta \int \frac{ \phi_x }{\phi_x+\beta} \d x
\ee
(index $ul$ is omitted in $\beta$ and $\delta$ as well as 
in the formulae below) 
is the probability per single interaction act that the line
photon will be absorbed by dust (obviously, $1-\delta$ is
the probability for a photon to excite a molecule).

If the source function does not vary much, we can
take it out from the integral (the so called on-the-spot or the first order
escape probability approximation) and get:
\beq \label{eq:J}
\overline{J} (\tau)& \simeq &(1-\delta)\left[ 1 - \frac{1}{2}K_2(\tau) -
 \frac{1}{2}K_2(2\tau_0-\tau) \right] S(\tau) \nonumber \\
&+& \delta \left[ 1 - \frac{1}{2}L_2(\tau) -
 \frac{1}{2}L_2(2\tau_0-\tau) \right]  B(\Td) \\
&=& (1-p) S+\pc B(\Td) \nonumber ,
\eeq
where
\beq \label{eq:KL2}
K_2(\tau)&=&2\int_\tau^\infty K_1(t) \d t \nonumber \\
& = &
\frac{1}{1-\delta } \int \frac{\phi_x^2 }{\phi_x+\beta}
E_2 \left( [ \beta +\phi_x ] \tau \right) \;  \d x , \\
L_2(\tau)&=&2\int_\tau^\infty L_1(t) \d t \nonumber \\
&=& \frac{\beta}{\delta} \int \frac{\phi_x}{\phi_x+\beta}
E_2 \left( [ \beta +\phi_x ] \tau \right) \;  \d x .
\eeq
For every column density of the slab we can take the intensity
in the middle plane $\overline{J}(\tau_0)$ as the representative one.
We perform all our calculations using equation (\ref{eq:J}) with
$\tau=\tau_0=\alw H$ (where $H$ is the slab
half-thickness), i.e taking\footnote{\citet{CW95} and \citet{WW97} 
have considered a semi-infinite slab with $\tau_0=\infty$ and used the escape terms 
$p=\delta+(1-\delta)K_2(\tau)$ and 
$\pc=p-K_2(\beta=0,\tau)$, while the correct ones in that case are
$p=\delta+(1-\delta)K_2(\tau)/2$ and $\pc=\delta[1-L_2(\tau)/2]$.}
\be  \label{eq:esc1}
p=\delta+(1-\delta)K_2(\tau), \quad \pc=\delta[1-L_2(\tau)].
\ee
If, however, the line is optically thick, the escape is negligible
and the  mean intensity is simply
\be \label{eq:J2}
\overline{J} \simeq (1-\delta)  S+\delta B(\Td).
\ee
This expression can be obtained directly from equation (\ref{eq:rte}) assuming
isotropic intensity of radiation (i.e. assuming $\d I_x/\d z=0$ and $I_x=J_x$),
solving for $J_x$,
and integrating it over the frequency with a  weight $\phi_x$.
If  most of the considered lines are optically thick, the populations
do not depend on the optical depth of the considered point but
are determined by the local conditions (such as hydrogen density, 
gas-to-dust density ratio, and dust and gas temperatures) only.

Following \citet{HM79},  we approximate $\delta(\beta)$ as
\be \label{eq:delapp}
\delta \simeq \frac{2 \beta}{1+2 \beta}
 \left[ \ln{\left( {\rm e}+\frac{1}{\sqrt{\pi}\beta} \right) } \right]^{1/2}
\ee
and  slightly modify the expression for $K_2(\tau)$:
\be\label{eq:k2}
K_2(\tau) \simeq  \frac{\exp(-\taud)}{1+(1-\delta)\sqrt{\taud} }
\ \frac{1}{1+\tauc [2 \pi \ln (2.13 + \tauc^2)]^{1/2}} ,
\ee
where $\tauc=\tau/\sqrt{\pi}$ is the optical depth in the line center and 
$\taud=\beta \tau$ is the optical depth for the dust absorption. 
The continuum escape coefficient $\pc=\delta[1-L_2(\tau)]$ is approximated as
\beq
\pc&\simeq & X \left( \delta - \frac{\exp(-\taud)}{1+\tauc [\pi \ln (1.25 + \tauc)]^{1/2}}
\right.  \nonumber \\
&\times & \left. \frac{\delta+\taud[1-\ln(1+1/\tau_*) ]}{1+\taud} \right) ,  \\
X&=& 1-0.095 \frac{y}{1+y^2}\ln (1+0.028/\beta) , \nonumber
\eeq
where  $y=(0.477\tauc)^{0.277}$, $\tau_*=\max(\taud,\tauc)$.
These expressions for $K_2$ and $\pc$ are accurate within $10\%$ for
any $\beta$ and~$\tau$.


The masing lines need a different treatment.
We neglect the dust influence in such transitions, i.e. assume $\beta=\delta=\pc=0$.
Using expression (\ref{eq:KL2}), one can show that at 
large maser optical depth $|\tau|\gg1$
\be \label{eq:k2as}
K_2(\tau)\sim \frac{\exp(|\tauc|)}{|\tauc| \sqrt{\pi \ln |\tauc|}}.
\ee
Then the line escape coefficient for $\tau<0$ can be represented as
\be \label{eq:esc2}
p(\tau)= \frac{\exp(|\tauc|)+1.18|\tauc|}{1+|\tauc|\ [\pi\ln(1+|\tauc|)]^{1/2}} .
\ee
This expression reproduces the exponential asymptotic
behavior at large $|\tauc|$ (\ref{eq:k2as}) as well as matches 
the asymptotic of $K_2$  at small $\tauc$ (see eq.~\ref{eq:k2}).

We consider different types of dust.
The amorphous ice absorption coefficient is computed from
the optical constants of  \citet{LG83}.
The crystalline ice absorption coefficient are based on the
data from \citet{Be69} and \citet{Wa84}.
The silicate and graphite data are from \citet{LD93}.

\subsection{Statistical balance equations}

The populations of $M$  levels can be determined from the
population balance equations which, in the stationary case, take the form:
\beq  \label{eq:balance}
\sum_{\low \ne \up} \nj \, \Wji & =& \ni\sum_{\low \ne \up} \Wij, \quad \up=1,2 \ldots , M-1, \\
\sum_{\up=1}^{M} \ni &=& 1,  \nonumber
\eeq
where $\Wij=\Cij+\Rij$ is the total rate of the $\up \rightarrow \low$
transition,
\be \label{eq:colrate}
\Cji^{\uparrow}    =  \frac{\gi}{\gj}  \Cij^{\downarrow} \exp
\left(- \frac{h \nuij}{kT}\right), \qquad
\Cij^{\downarrow}  =  \Nh \kij(T)
\ee
being the rates of the collisional excitation and deexcitation of
water molecules by molecular hydrogen which kinetic temperature
equals that of the water. 
The rates of the radiative excitation  and deexcitation are given by
\be  \label{eq:radrate}
\Rji^{\uparrow}    =  \Aij \frac{\gi}{\gj} \Jij , \qquad
\Rij^{\downarrow}  =  \Aij \left( 1+\Jij \right) .
\ee
The Einstein A-coefficients are taken from \citet{CVK84}.
Collisional deexcitation rates  are from \citet{Gr93}, while the
excitation rates are computed using the detailed balance condition.
We take into account the first 45 rotational levels of ortho- and para-\water\ molecules
and all possible radiative and collisional transitions between them.

Substituting the radiation intensity from (\ref{eq:J}) into
(\ref{eq:radrate}) and the statistical balance equations (\ref{eq:balance}),
we obtain:
\beq \label{eq:balafinal}
& &\sum_{\low > \up} \Aji \left( \pji \nj +\pdji \Bd_{\ji}
\gj \Delta \nij \right)  \nonumber \\
& - & \sum_{\low<\up} \Aij \left(  \pij \ni + \pdij \Bd_{\ij}
\gi \Delta \nji  \right)        \\
& = & \sum_{\low \ne \up} \left(  \Cij  \ni -  \Cji  \nj \right)   , \quad u=1,2,\dots,M-1,
  \nonumber
\eeq
with the same normalization as before $\sum_{\up} \ni=1$.


The populations can be found from the solution of the system of non-linear algebraic
equations  (\ref{eq:esc1}), (\ref{eq:esc2}), and (\ref{eq:balafinal}).
We use the Newton-Raphson method with line searches and backtracking
\citep{NR92}. We normally start from high hydrogen density where
the solution is close to the Boltzmann at the gas temperature and use the
obtained solution as a zeroth approximation for the next point.
We continue iterations until the maximum error in the system becomes
smaller than $10^{-11}$ and the maximal change in populations is smaller than
$10^{-9}$. It takes 5 to 15 iterations for most sets of parameters and
each computation takes on average
about 15 ms CPU time on a Pentium IV 2 GHz Linux PC.

\subsection{Maser intensity} 

If the  maser amplification takes in the homogeneous
medium and there is no velocity gradients, 
the maser intensity can be estimated as 
\begin{equation} \label{eq:masint}
I_x(\taum) \simeq (I_0+|S_{\rm m}|) \exp(\taum\phi_x),
\end{equation}
where $I_0$ is the background continuum intensity,
$S_{\rm m}$ is the source function for the $\up\rightarrow \low$   masing transition,
and  the optical depth  in the maser line $\taum$ is
\begin{equation} \label{eq:tau}
\taum=\alpha_{\rm m}L_{\rm coh}=
\frac{\lambdaij^3 \gi \Aij}{8 \pi c (\Delnud/\nuij) } \delnm \Nw L_{\rm coh},
\end{equation}
where 
$\delnm=\nup/\gu-\nlow/\gl$ (positive for the maser) and
$L_{\rm coh}$ is the typical coherent length of the maser amplification.
If the excitation temperature of the masing transition is larger
than the brightness temperature of the background, one can neglect $I_0$
(and obviously other way around too). 
We assumed the first possibility (i.e. $I_0=0$).
In this paper we quote values of $\taum$ for 
$L_{\rm coh}=H$. Obviously, at grazing angles to the slab surface 
this value can be exceeded many times.

\subsection{Parameter range}

The water level populations depend on the hydrogen
density $\Nh$, the dust and water concentrations and
temperatures as well as the slab half-thickness $H$.
When the photon escape from the surface is negligible because
the slab is optically thick either in continuum ($\taud\gtrsim 1$) or
in the line ($\tau\gg1$), the terms $K_2$ and $L_2$ can be
omitted in the escape probabilities (\ref{eq:esc1}) (i.e. $p=\pc=\delta$).
The solution of the population balance equations then depends
on $\delta$, which in turn depends only on the ratio of
the absorption coefficients $\beta$, but not
on the absorption coefficients individually. Since
the dust opacity $\kabs$ has a rather
weak dependence on the size of the dust particles, it
is rather natural to use
the water-to-dust mass ratio $\fwd$ as a parameter,
where $\fw$ is the water-to-hydrogen mass ratio and
$\fd$ is the dust-to-hydrogen mass ratio. 
The water concentraion relates then to these parameters 
through $\Nw=\Nh(\fwd) \fd/9$ (factor 9 comes from the 
ratio $m_{\rm H_2O}/m_{\rm H_2}$). 
When the escape through the surface is small  and masers are unsaturated
(i.e. a masing transition does not affect the level populations),
the populations do not depend on $\fd$ and 
the scale-height $H$, and the optical depth is then
linearly proportional to $\fd H$.
In such a situation the main parameters are
$\Nh$, $\fwd$, $T$ and $\Td$.  If the dust is suffiently cold, its own
radiation is then negligible and dependence on $\Td$ disappears.
We normally consider the dust colder than the gas, $\Delta T\equiv T-\Td>0$,
because this is required for inversion to occur (in the optically thick case).

One can also point out that for large $\fwd$ when $\beta\sim\delta\ll 1$
in the main infrared pumping lines,
the solution of the balance equations (\ref{eq:balafinal})
depends on the product $\Nh\fwd$, but not on $\Nh$ and $\fwd$ individually,
because $\Cij/\pij\propto \Nh/\beta \propto \Nh \fwd$.

We consider a range  of the hydrogen  concentrations from
$10^8$ to $10^{12.5}\cminv3$. At higher  $\Nh$, the inversion is absent
because of the thermalization by collisions with the hydrogen molecules, while
at smaller $\Nh$ we can extrapolate the results from $\Nh=10^8$.

In astrophysical objects it is easier to estimate  the dust-to-gas
mass ratio $\fd$ than other parameters of interest. We
fix it at a standard for the interstellar medium value of $\fd=0.01$.
This then determines the dust optical depth 
\be 
\taud=0.033 \kabs \frac{\Nh}{10^{10}} 
\frac{\fd}{0.01} \frac{H}{10^{14}}. 
\ee

For the majority of calculations we consider the dust in the form
of the amorphous ice grains  of the size $a=0.1\ \mum$. 
Comparison is also made to different sizes and temperatures of dust 
as well as other types of dust (crystalline ice, silicate, and graphite). 

The water-to-dust mass ratio probably varies by orders of magnitude from
one object to another. Therefore, we consider a broad range of
$\fwd$ from $10^{-8}$ to $10^{0}$.
The maximum possible $\fw$ in the interstellar medium
is about $10^{-3}$ which transforms to the maximum $\fwd\sim0.1$
for  $\fd=0.01$.
However, for a smaller dust content, $\fwd$ can be larger.

\begin{figure}
\centerline{\epsfig{file=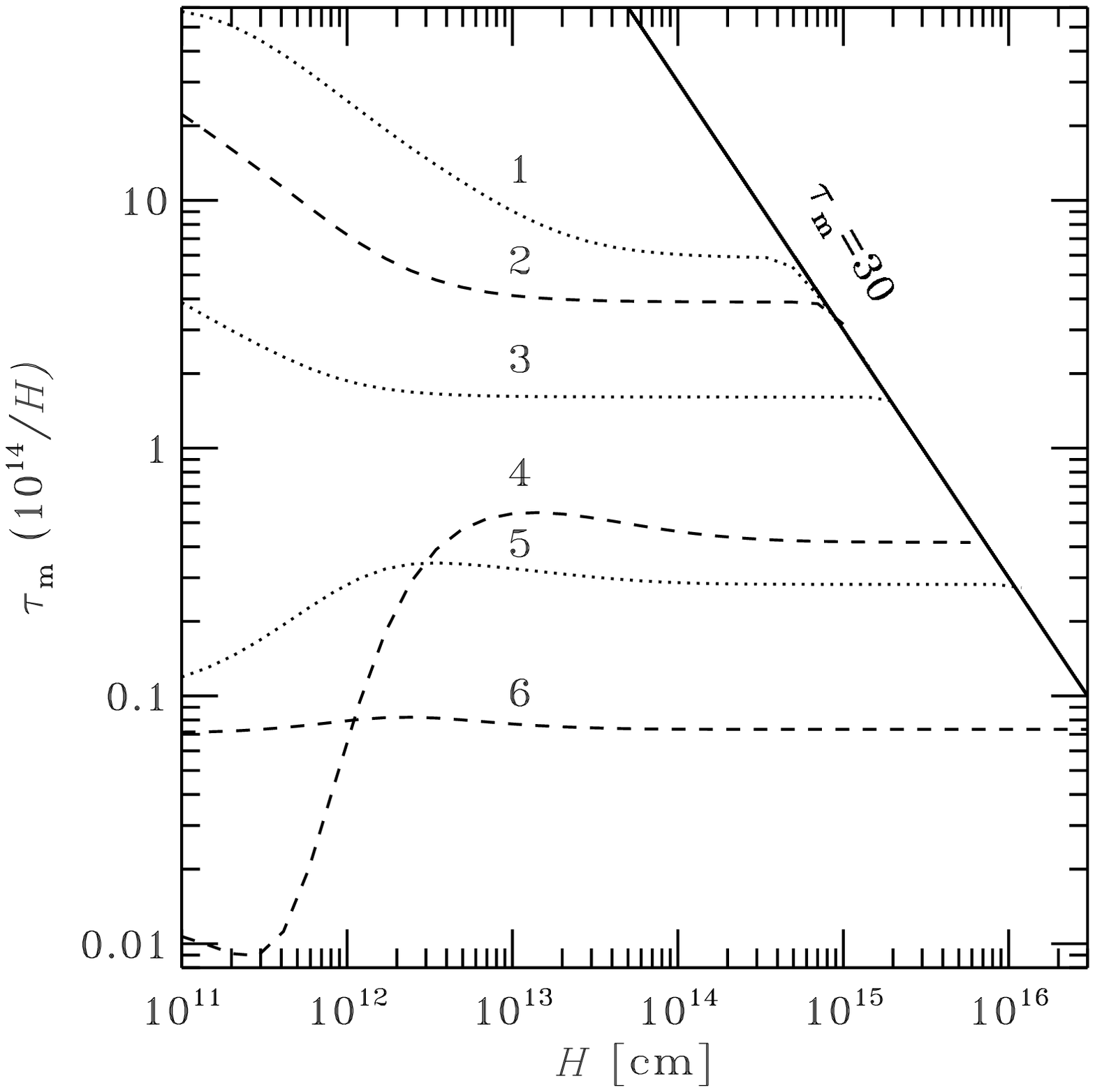,width=7.5cm}}
\caption{
Dependence of the $6_{16}$--$5_{23}$
maser absorption coefficient $\alpha_{\rm m}/(10^{-14}\cminvone)$
on the half-thickness of the slab
for six sets of parameters $(\Nh,\fwd)$: (1) $10^{9},10^{-2}$;
(2) $10^{10},10^{-3}$; (3) $10^{11},10^{-4}$;
(4) $10^9,10^{-4}$; (5) $10^{10},10^{-5}$; (6) $10^{11},10^{-6}$.
Other parameters are fixed at $\Td=130$ K, $T=250$ K, $\fd=0.01$.
At a slab thickness less that $10^{13}$ cm, the effect of the boundary
becomes significant. For higher $H$, the maser becomes saturated when $\taum=30$.
}
\label{fig:tauH}
\end{figure}

Figure~\ref{fig:tauH}  shows the absolute value of the absorption coeffient
(in units of $10^{-14}\cminvone$)
in the main masing line $6_{16}\rightarrow 5_{23}$  for six representative
sets of parameters $(\Nh,\fwd)$.
One sees that at $H\lesssim 10^{13}$ cm, the surface
escape of photons (the de Jong mechanism) starts to influence
the populations.
At high $H$, the optical depth $\taum$ can be sufficiently large  to saturate  the maser.
One can easily estimate the optical depth when it happens. 
The saturation occurs when the induced transition rate $A\overline{J}$ 
in the masing line becomes comparable to the rates of transitions 
from the upper masing level in other strong lines
(see eq.~\ref{eq:radrate}) which are $\Rij\sim 1$. 
Because the Einstein $A$ coefficient in the 1.35 cm $6_{16} \rightarrow 5_{23}$
masing line is about $10^{-9}$
(and about $10^{-5}$ in other masing transitions), the saturation
occurs when $\overline{J}\sim 10^9$. On the other hand, since 
\be 
\overline{J}\sim p|S|\sim \exp(\taum/\sqrt{\pi})|S| 
\ee
(see eq.~\ref{eq:esc2}) and $|S|\sim 30$,  
the maser saturates at the optical depth $\taum\sim 30$ 
(which weakly depends on the value of $S$). 
This limiting optical depth
is reached at different $H$ depending on the conditions.
Here the maser saturates and the inversion decreases.

The inversion is rather flat around $H\sim 10^{14}$ cm, 
where it is a function of the local conditions only.
Therefore, we use this height for our calculations.
We should also note that our results do not depend on the geometry of
the system, and can be applied
not only to the slab but to any other geometry.

\begin{figure}
\centerline{\epsfig{file=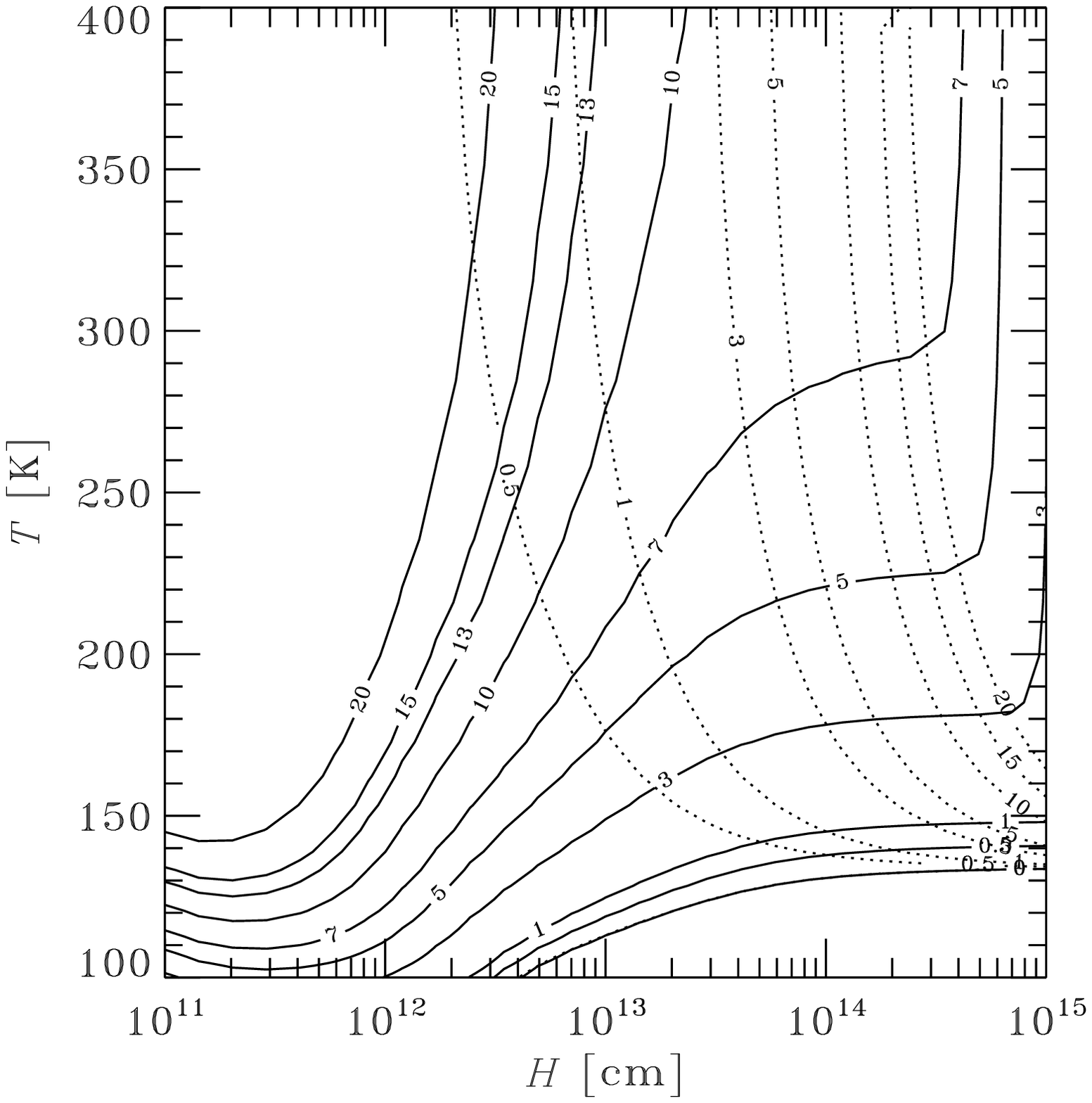,width=7.5cm}}
\caption{
Levels of the constant maser absorption coefficient $\alpha_{\rm m}/(10^{-14}\cminvone)$
(solid curves) and the maser optical depth $\taum$ (dotted curves) on the plane
slab half-thickness -- gas temperature
for $\Nh=10^{9}\cminv3$ and $\fwd=10^{-2}$.
Other parameters are fixed at $\Td=130$ K, $\fd=0.01$.
}
\label{fig:TH}
\end{figure}

The contours of the constant maser absorption coefficient
and the optical depth on a plane $H$ - gas temperature $T$
are shown in Fig.~\ref{fig:TH}. One sees that, independently of the
gas temperature, the absorption coefficient is rather flat around $H\sim 10^{14}$ cm.
The flat part becomes shorter at  higher gas temperature, because
the saturation ($\taum\sim 30$) happens for smaller $H$.
One should note here that the parameter set $\Nh,\fwd$ taken at this
graph gives much shorter flat parts than other sets presented in Fig.~\ref{fig:tauH}.
Our fiducial $H=10^{14}$ cm still seems to be a good choice for studying in details
the physics of the unsaturated water maser in a dusty environment.
It is also interesting to note that for small $H$ inversion exist even
when $T<\Td$ because of the action of the de Jong mechanism.

\section{Analytical four level model}
\label{sec:anal}

Our main goal is to study in details the inversion mechanism
of the $6_{16}\rightarrow 5_{23}$  masing transition.
Before we proceed to the results of calculations, it is useful
to understand the physical processes responsible for the action of this main
maser using a simplified system of molecular levels, which mainly participate
in the maser pumping:  $4_{14}$ (level 1), $5_{05}$ (level 2),
$5_{23}$ (level 3) and $6_{16}$ (level 4) (see Fig.~\ref{fig_pump} and
Deguchi 1981).
The upper maser level 4 interacts
mostly with the 2nd level which in turn interacts with level 1,
while the lower maser level 3 interacts directly mostly with level 1.
(We neglect here rather strong transitions from the 2nd to the 3rd level via
level $5_{14}$.)
The rates of radiative (in the case of the unsaturated maser)
as well as collisional transitions via the masing line
$4\rightarrow3$  are much smaller than the rate of transitions to other levels.
Thus, we can assume that the populations at the masing levels
are completely unrelated to each other, and depend only on the
rate of the exchange to other levels.
This allows us to write a system of three (two-level) population balance equations
that relate the corresponding populations in the standard form:
\be
\nup(\Rul+\Cul)=\nlow(\Rlu+\Clu),
\ee
where $\ul$ are $21$, $42$ and $31$.
From equation~(\ref{eq:sour}), we get
the corresponding (dimensionless) source functions:
\be
S=\frac{1}{1+\epsilon}\J+\frac{\epsilon}{1+\epsilon}\Bg,
\ee
where $\epsilon=C(1-\exp[-E/T])/A$, $E$ is the corresponding transition
energy in Kelvin and we omitted the indices $ul$.
Assuming that the photon escape is negligible, we
substitute the expression for the radiation field (\ref{eq:J2}) and 
obtain:
\be  \label{eq:sbb}
S=\frac{\xi}{1+\xi}B+\frac{1}{1+\xi}\Bd ,
\ee
where $\xi\equiv\epsilon/\delta$.
The intensity is then
\be
\J=\frac{\xi-\epsilon}{1+\xi}B+
\frac{1+\epsilon}{1+\xi}\Bd .
\ee
This is, of course, not a self-consistent solution, because $\delta$
depends on $\beta$ which  in its turn is a function of the level populations.
We can assume in the first approximation (for calculating $\beta$)
that there is a Boltzmann distribution of the populations at the gas temperature $T$.
We thus see that the source function is determined  by a single parameter, the
ratio $\xi$. We also can note that for a large water content (or small dust content,
i.e. $\beta\ll 1$) $\delta\sim\beta\propto 1/A$ and $\xi$ does
not depend on the line strength (since $\aul$ cancels out). Then $S$ 
is completely determined by the relative ratio of the photon destruction rate
by collisions to that by dust absorption.
On the other hand, when the dust is dominating the absorption (i.e. $\beta\gg 1$),
$\delta\sim 1$ and $\xi=\epsilon$, the source function is determined
just by the relative importance of collisional and spontaneous
radiative transitions.
Now we can obtain the general expression for the
inversion in the levels 4-3. Since
$\nup/\gu=\nlow/\gl\ \Sul/(1+\Sul)$,
the inversion is
\be \label{eq:inv}
\delnm=\frac{n_1}{g_1}\left[ \frac{S_{21}}{1+S_{21}}
\frac{S_{42}}{1+S_{42}} - \frac{S_{31}}{1+S_{31}}\right].
\ee
Because the $1\leftrightarrow3$ transition is much more affected by
dust that other main transitions, 
it is possible that the population of the lower masing level is determined by
the dust temperature (i.e. $S_{31}=\Bd_{31}$),
 while in other transitions the dust influence is still negligible
(i.e. $S_{42}=\Bg_{42}$, and $S_{21}=\Bg_{21}$).
The inversion then reaches the maximum
\be \label{eq:genii} 
\delnm= \frac{n_1}{g_1}\left( {\rm e}^{-E_{41}/T} -
{\rm e}^{-E_{31}/\Td} \right) .
\ee
Let us now consider some limiting cases.

\subsection{Large water content}
\label{sec:smalldust}

When the water concentration is large, 
the source functions depend on one parameter $\xi=\epsilon/\delta$. 
This means that the inversion is the function of $\Nh\fwd$ only.
If $\xi$ for all main lines is very small
(e.g. small hydrogen density) then the levels are thermalized at the dust temperature,
while for large $\xi$, thermalization occurs at the gas temperature.
The limits on $\Nh$ where the inversion is possible depend on $\fwd$.

Let us first investigate the limit $\xi\ll1$.
The source functions can be represented then as
$S=\Bd+\xi \Delta B$, where $\Delta B\equiv\Bg-\Bd$.
Then
\be
\frac{S}{1+S}=\exp(-E/\Td) \left[ 1+\xi \frac{\Delta B}{\Bd(1+\Bd)}\right].
\ee
For the amorphous ice the $\xi$ parameters can be approximated as 
\beq
\xi_{21}&\approx & 6\times 10^{-5} \Nh \fwd, \nonumber \\
\xi_{42}&\approx&  10^{-5} \Nh \fwd,  \\
\xi_{31}&\approx& 10^{-7} \Nh \fwd. \nonumber 
\eeq
Because, 
$\xi_{31}\ll \xi_{42}\ll \xi_{21}$, we can keep only the terms
with $\xi_{21}$ in equation (\ref{eq:inv}). 
A condition of the positive inversion $\delnm>0$ then transforms into
\be  \label{eq:xi12min}
\xi_{21}>\xi_{21,\min}=\frac{\eem}{\Td}\frac{\Bd_{21}(1+\Bd_{21})}{\Delta B_{21}},
\ee
which corresponds to $\Nh\fwd>120$ for  $\Td=130$ K. 
Here $\eem=E_{41}-E_{31}\approx 1$ K is the maser transition energy. 
For smaller amount of water, the inversion disappears.
However, we implicitly assumed here that $\xi\Delta B$ is a small correction
to $\Bd$. If the dust is sufficiently cold, this is not true anymore.
Thus, if $\Bd\ll \xi \Bg\ll 1$, the source function is just $S\approx\xi\Bg$.
Because $\xi_{21}$ is by far the largest among the considered
transitions, the inversion appears when
\be
\xi_{21} >\xi^*_{21,\min}=\frac{\xi_{31}}{\xi_{42}}
\frac{\Bg_{31}}{\Bg_{21}\Bg_{42}},
\ee
corresponding to $\Nh\fwd>50$. Thus generally the inversion exists at
$\xi_{21}>\max[\xi_{21,\min},\xi^*_{21,\min}]$.

In another extreme case, small influence of the dust relative to collisions,
i.e. $\xi\gg 1$, we get
$S=\Bg-\Delta B/\xi$. Now keeping only the term $\propto 1/\xi_{31}$
we obtain a condition for the inversion
\be \label{eq:xi13max}
\xi_{31}<\xi_{31,\max}=\frac{T}{\eem}\frac{\Delta B_{31}}{B_{31}(1+B_{31})}, 
\ee
which corresponds to $\Nh\fwd<10^9$ for $T=250$ K.

We can get an estimate for the temperature difference
needed to produce the inversion.
The inversion appears first in the region where the $3\rightarrow1$ transition
is dominated by dust (i.e. $\xi_{31}\ll 1$ and
$S=\Bd+\xi \Delta B$),
while for other main transitions the dust influence
is still negligible (i.e. $\xi_{21}\gg\xi_{42}\gg 1$ and
$S=\Bg-\Delta B/\xi$).
Expanding $\Delta B$ in the vicinity of $\Td$ as
 $\Delta B=(\Bd/\Td)^2 E \Delta T$,
we get from the condition $\delnm>0$,
\be \label{eq:tminsmall}
\Delta T > \frac{\eem\Td}{E_{31}\left( 1-\xi_{31}{\rm e}^{-E_{31}/\Td}
\right)-E_{42}
{\rm e}^{-E_{42}/\Td} /\xi_{42}},
\ee
where we neglected the terms of the order $1/\xi_{21}$.
For $\Nh\fwd=10^6$, we have $\xi_{31}=0.1$ and $\xi_{42}=10$, 
and then $\Delta T > 0.5$ K for $\Td=130$ K. 
In the limit $\xi_{31}\ll 1$ and $\xi_{42}\gg 1$,
corresponding to
$S_{42}=B_{42}$, $S_{21}=B_{21}$ and $S_{31}=\Bd_{31}$,
we get a much simpler expression
\be \label{eq:tminsmall2}
\Delta T  > \eem \Td/E_{31} = 0.003\Td.
\ee
We see that already a small difference in the temperatures  produces the
inversion.

\subsection{Small water content}
\label{sec:largedust}

A small water content (or large dust content) is described by
$\beta\gg 1$ and $\delta\sim1$.
For the slab thickness $H=10^{14}$ cm, 
the medium is transparent for small $\fwd$ and small $\Nh$ 
(since the dust optical depth is proportional to the
hydrogen density), and thus a similar analytical
description as above is not valid. However, for larger $H$, and/or $\Nh$,
and/or $\fd$, the medium still can be sufficiently opaque.  
Then, the source functions in all pumping lines are
\be \label{eq:smallS} 
S=\frac{\epsilon}{1+\epsilon}\Bg+\frac{1}{1+\epsilon}\Bd .
\ee
Thus the inversion does not depend anymore on $\fwd$, but 
is defined only by the hydrogen density
(which determines $\epsilon$) and the temperatures.
At very high $\Nh$, $\epsilon\gg 1$ and all source functions
are given by the Planck function at the gas temperature,
while for small $\Nh$, $\epsilon\ll 1$ and thermalization occurs at
the dust temperature.
We can determine the condition when the inversion appears
from the equation $\delnm=0$.
For small $\Nh$, all $\epsilon\ll1$ and the
source functions for every transition can be written as
$S= \Bd+\epsilon \Delta B$.
Substituting this into equation (\ref{eq:inv}) and
keeping only the terms of the first order in $\epsilon$,
one gets the lower limit on the density where the inversion is
still possible:
\beq
\Nh^{\min}&=&\frac{\eem}{\Td}\left\{
\frac{\Delta B_{21}(\epsilon_{21}/\Nh) }{\Bd_{21}(1+\Bd_{21})} \right. \\
&+& \left. \frac{\Delta B_{42}(\epsilon_{42}/\Nh) }{\Bd_{42}(1+\Bd_{42})} -
\frac{\Delta B_{31}(\epsilon_{31}/\Nh) }{\Bd_{31}(1+\Bd_{31})}
\right\}^{-1}   \nonumber
 \eeq
(note that $\epsilon$ is proportional to the
hydrogen density).
For example, for $T=250$ K and $\Td=130$ K, we get limiting
 $\Nh^{\min}=2\times 10^8\cminv3$.

Analogously, we can get an upper limit on $\Nh$ assuming
that $\epsilon\gg 1$. Rewriting  $S=\Bg-\Delta B/\epsilon$ and
keeping the terms of the first order in $1/\epsilon$ we get
\beq\label{eq:hren} 
\Nh^{\max}&=&\frac{T}{\eem}\left\{
\frac{\Delta B_{31}(\Nh/\epsilon_{31})}{B_{31}(1+B_{31})} \right. \\
&-&\left. \frac{\Delta B_{21} (\Nh/\epsilon_{21})}{B_{21}(1+B_{21})} -
\frac{ \Delta B_{42}(\Nh/\epsilon_{42})}{B_{42}(1+B_{42})}
\right\}  .  \nonumber
\eeq
For the same parameters as above,  this expression gives
$\Nh^{\max}\sim 10^{13}\cminv3$.

\begin{figure*}[htb]
\centerline{\epsfig{file=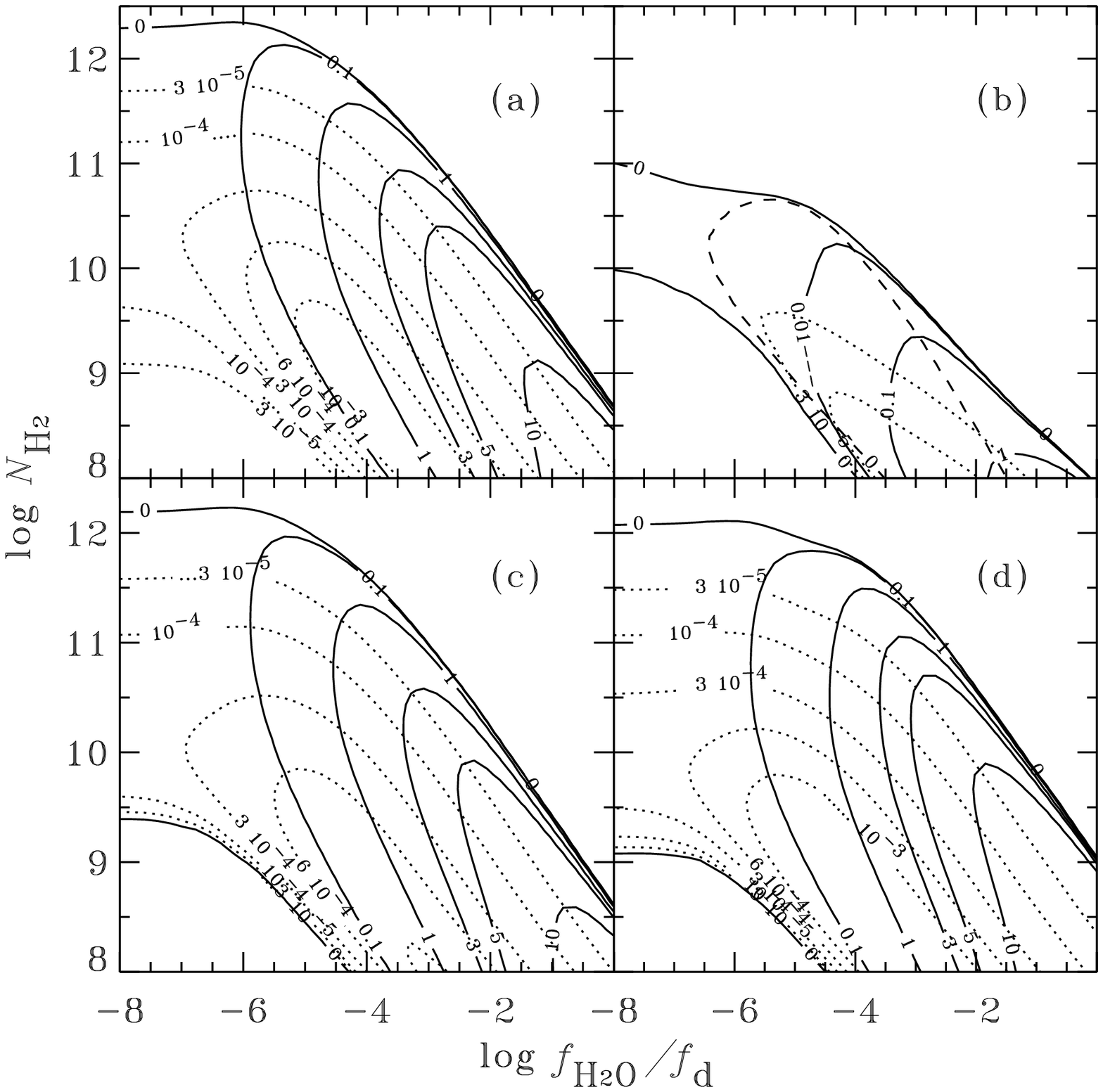,width=16cm}}
\caption{
Contour plots of the levels of the constant inversion $\delnm$ (dotted curves) and
constant optical depth (solid curves)
for the  $6_{16}$--$5_{23}$ transition at the plane
hydrogen concentration --  gas-to-dust concentrations ratio.
A slab of the half-thickness $H=10^{14}$ cm and the dust in the form of the
amorphous ice of the size $a=0.1\mum$ was assumed with the dust-to-gas
mass ratio $\fd=0.01$.  The panels correspond to the following temperatures
(a) $\Td=30$ K, $T=250$ K; (b)   $\Td=130$ K, $T=132$ K; 
(c) $\Td=130$ K, $T=250$ K; (d) $\Td=130$ K, $T=400$ K. 
The dashed contour in panel (b) corresponds to the negligible photon escape case 
$p=\pc=\delta$.
}
\label{fig:4panels}
\end{figure*}

Now let us obtain a minimum temperature difference needed for the inversion.
Representing the source function as $S=\Bd+\displaystyle
\frac{\epsilon}{1+\epsilon} \Delta B$
and expanding $\Delta B$ for small $\Delta T$,
we get
\beq \label{eq:deltmin}
\Delta T_{\min}&=&\eem \Td  \left\{
\frac{\epsilon_{21} E_{21}}{1+\epsilon_{21}}  {\rm e}^{-E_{21}/\Td} \right. \\
&+& \left. \frac{\epsilon_{42} E_{42}}{1+\epsilon_{42}}{\rm e}^{-E_{42}/\Td}
-\frac{\epsilon_{31} E_{31}}{1+\epsilon_{31}}  {\rm e}^{-E_{31}/\Td}
\right\}^{-1}  . \nonumber
\eeq
For our fiducial $\Td=130$ K, substituting  $\Nh=3\times 10^{10}\cminv3$
 we get $\Delta T_{\min}=3.5$ K.

\begin{figure}[htb]
\centerline{\epsfig{file=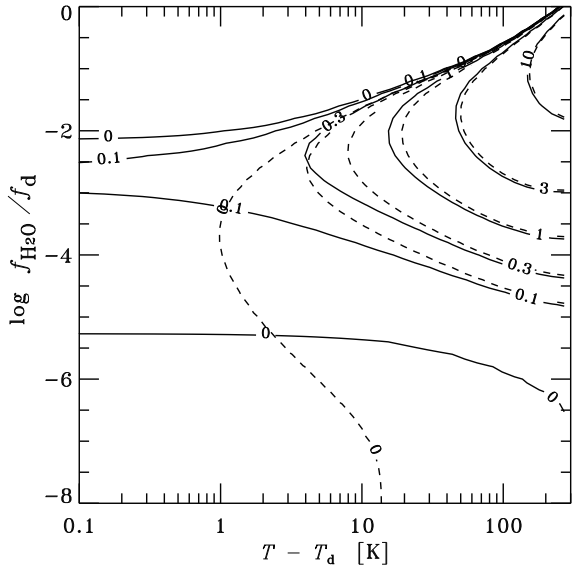,width=7.5cm}}
\caption{
Contour plots of the levels of the constant $6_{16}$--$5_{23}$ 
maser optical depth $\taum$  (solid curves) at the plane
gas-dust temperature difference -- gas-to-dust mass ratio  for the
following set of parameters  $\Nh=10^{9}\cminv3$,
$\Td=130$ K, $\fd=0.01$, and $H=10^{14}$ cm.
The dashed curves show the corresponding contours 
if the de Jong mechanism is not operating (i.e. the photon
escape from the slab is neglected and the $K_2$ and $L_2$
terms in equation (\ref{eq:esc1}) are omitted).}
\label{fig:s3}
\end{figure}

\subsection{Comparison to the de Jong model}
\label{dejong}

Let us compare the de Jong model to the cold dust model.
Using the escape probability approximation,
the radiation field is now $\J=(1-p)S$, where
$p$ is the probability for photons to escape without interactions and
we neglected here the influence  by the dust. The source function is then
\be
S=\frac{\epsilon}{\epsilon+p} \Bg.
\ee
In the optically thin case,
$p\rightarrow1$, and the source function is then identical
to that in the case of the large amount of the cold dust ($\Bd=0$, see
eq.~[\ref{eq:smallS}]).
Thus, in the optically thin limit the de Jong model is
similar to the (very) cold dust -- hot gas model.

\section{Results}    
\label{sec:result}

\subsection{Maser transition $6_{16}$--$5_{23}$}

\begin{figure}
\centerline{\epsfig{file=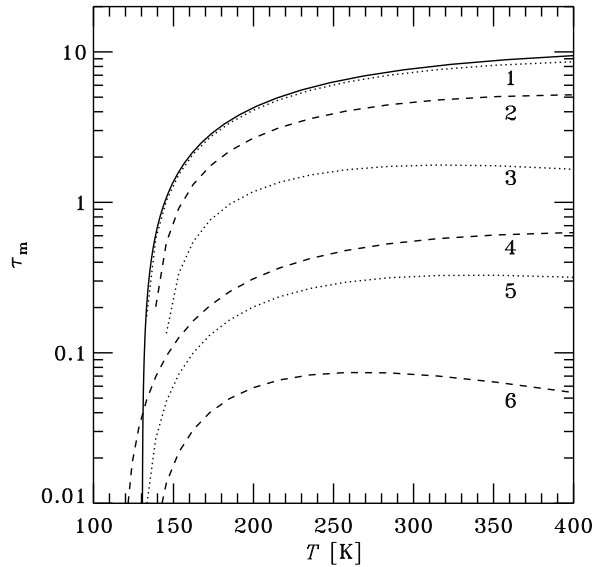,width=7.5cm}}
\caption{
Temperature dependence of the maser optical depth for
the same  six sets of parameters $(\Nh,\fwd)$ as in Fig.~\ref{fig:tauH}.
We fixed $H=10^{14}$ cm, $\Td=130$ K, $\fd=0.01$.
The solid curve corresponds to analytical formula (\ref{eq:tautemp}) with 
free normalization.
}
\label{fig:s2}
\end{figure}

The main results of our calculations for the case of amorphous ice 
of the size of $0.1\ \mum$ are shown on  Figs.~\ref{fig:4panels}-\ref{fig:s2}.
The dependences of the inversion and the maser optical depth
on $\Nh$ and $\fwd$ for the fixed 
dust and gas temperatures are shown in Fig.~\ref{fig:4panels}.
One sees that when the temperature difference is small (Fig.~\ref{fig:4panels}b),
the inversion appears in the region where
$\Nh\fw/\fd\sim10^{6}$ within two orders of magnitude.
The inversion is weak for small $\fwd$. 
We tested the influence of the de Jong mechanism on the results 
in this region by neglecting the terms  $K_2$ and $L_2$ in the escape 
probabilities (\ref{eq:esc1}). In that case (dashed curve in Fig.~\ref{fig:4panels}b),
the inversion  is absent there because a higher 
temperature difference is needed (see eq.~[\ref{eq:deltmin}]) to 
invert the populations. 

When $\Delta T$ is larger (see Fig.~\ref{fig:4panels}c,d),
the inversion becomes larger and the inversion region increases in size,
spreading also over to the region with a small water content ($\fwd<10^{-6}$).
Here the radiation field  is  completely dominated by the dust
and the inversion depends on the collisional rates defined by $\Nh$ only 
(for the fixed $T$).
At lower dust temperature $\Td=30$ K
(Fig.~\ref{fig:4panels}a), the optical depth
becomes larger by about 20\% comparing with $\Td=130$ K (Fig.~\ref{fig:4panels}c). 

We should point out that the inversion disappears at $\Nh\sim10^{12}\cminv3$
because of the level thermalization by collisions. This is lower than 
our analytical estimate (\ref{eq:hren}) which neglected many collisional
transitions.  On the right side the inversion region is bounded by 
$\Nh\fwd<10^{8.5}$ for $T=250$ K, which is very close to our 
estimate (\ref{eq:xi13max}).

Let us note that the maximum inversion occurs when
the water-to-dust mass ratio $\fwd\sim 10^5/\Nh$.
The maser optical depth, on the other hand, is proportional
to $\Nw \delnm$
which reaches the maximum at  $\fw/\fd\sim 10^{7.5}/\Nh$ for $T=250$ K.
This corresponds to $\Nw=10^{6.5}\fd$. The inversion disappears at
$\Nw>10^{7.5}\fd$ because the amount of dust (comparing with water) 
is not sufficient here to produce the inversion. 
Thus for a given dust content there exists
an optimal water concentration that produces the strongest maser.
In order to calculate the maser intensity (see eq.~[\ref{eq:masint}])
or the brightness temperature in the line center 
$T_{\rm br}\approx |T_{\rm ex}|\exp(\taum/\sqrt{\pi})$, 
one needs to have a rough estimate for  the excitation temperature.
Because the dependence is linear, the error in that is not important. 
At the maximum of $\taum$, the excitation temperature varies 
between $-10$ and $-100$ K when $T$ varies between 150 and 500 K.  

\begin{figure*}[htb]
\centerline{\epsfig{file=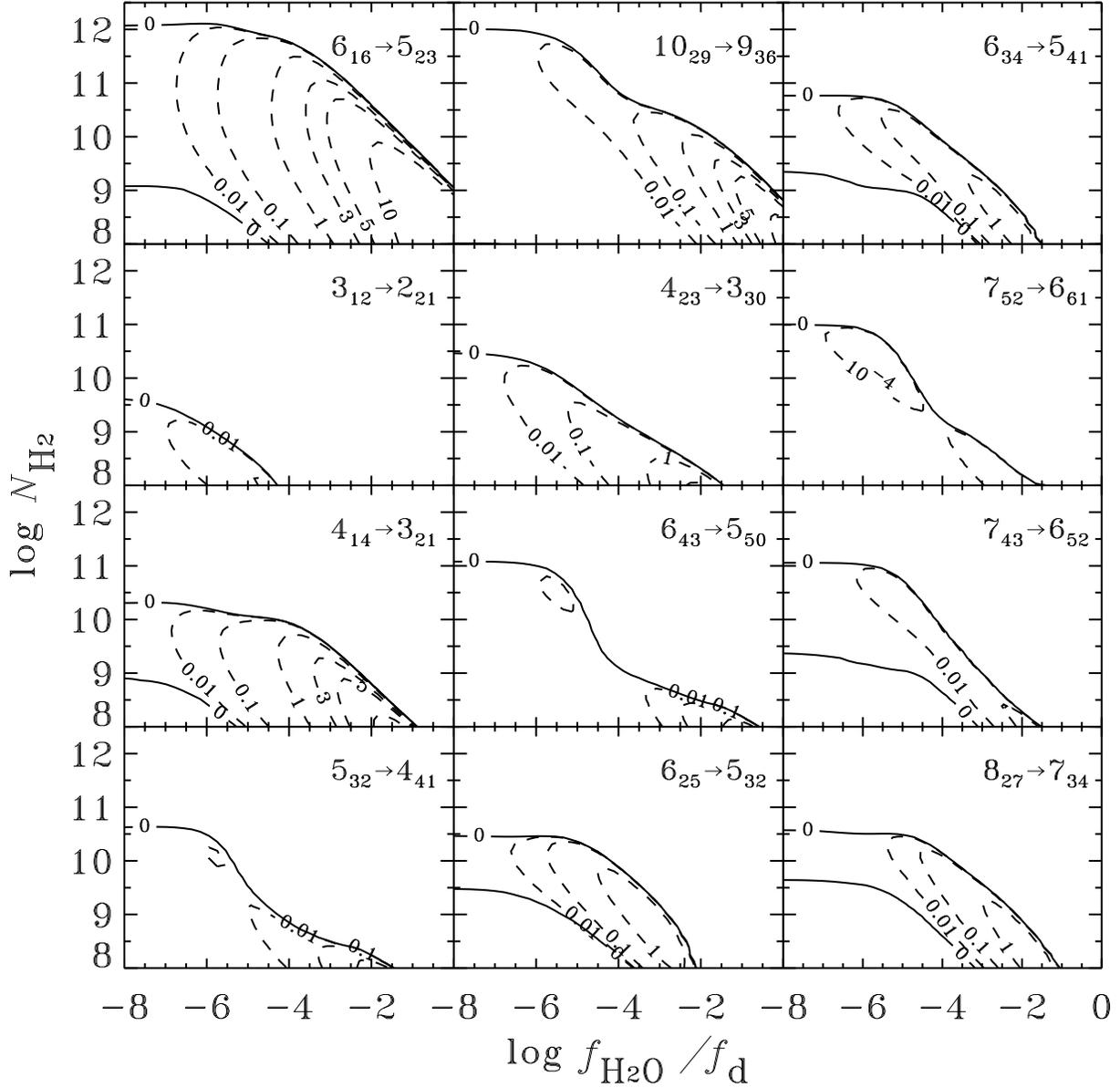,width=16cm}}
\caption{
Contours of the constant maser optical depth for
the strongest masing  transitions of ortho-water
for $\Td=130$ K, $T=400$ K.
Other parameters are the same as in  Fig.~\ref{fig:4panels}.
}
\label{fig:allmas}
\end{figure*}

\begin{figure*}[htb]
\centerline{\epsfig{file=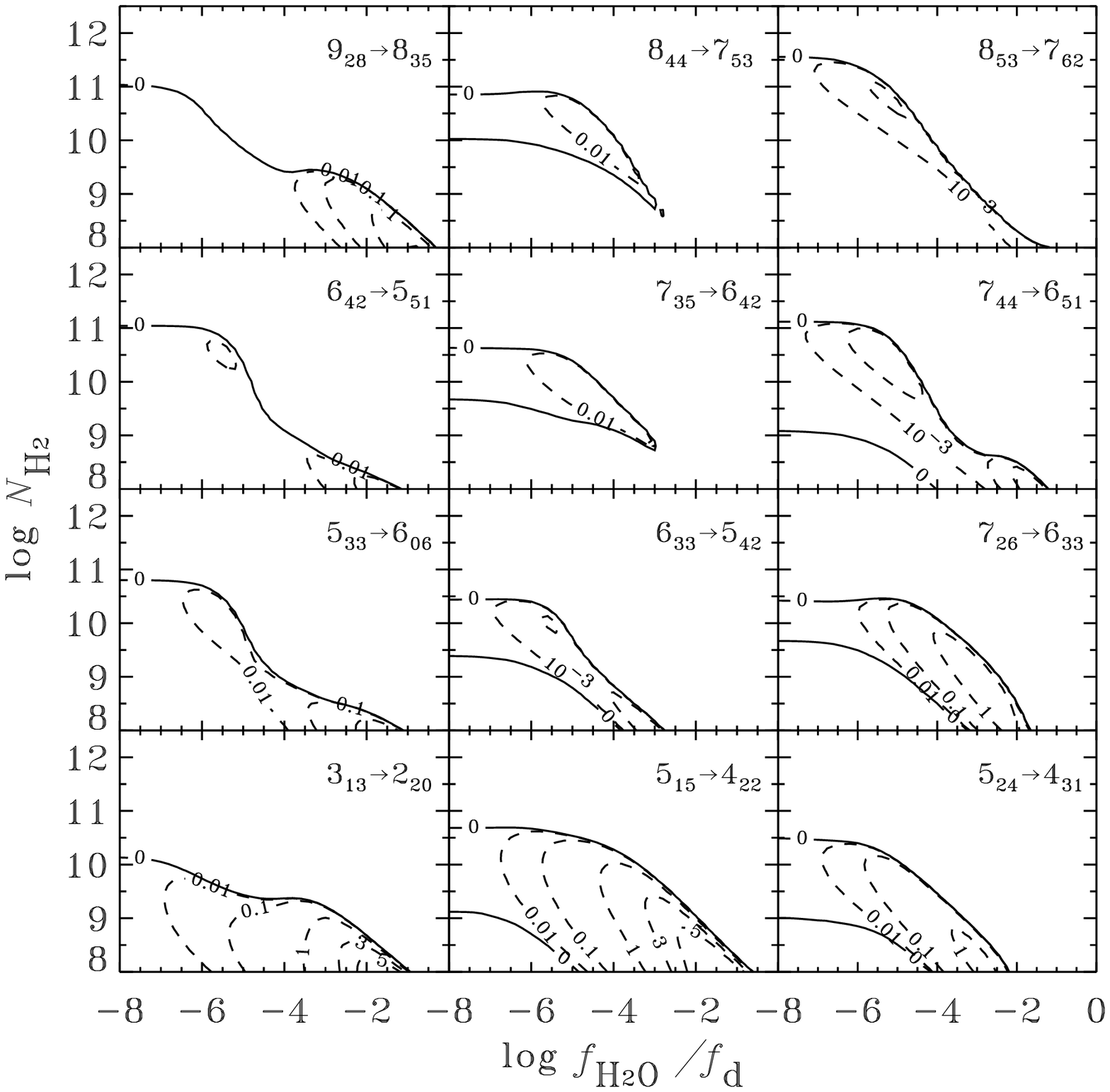,width=16cm}}
\caption{
Same as in Fig.~\ref{fig:allmas}, but for para-water.
}
\label{fig:allpara}
\end{figure*}

On  Fig.~\ref{fig:s3} we present the dependence of $\taum$ 
on $\fwd$ and $\Delta T$ for the fixed
value of $\Nh=10^{9}\cminv3$. We see that the inversion
disappears at large $\fwd$
because the levels are then thermalized at the gas temperature.
At small $\fwd$, the medium is optically thin for the line and 
the dust emission ($\taud<1$ for most infrared lines 
at this  $\Nh$ for $\fd=0.01$).
Here we also tested the influence of the photon escape from the 
surface. 
The dashed curves show the maser optical depth when the escape 
is neglected.
These results show that at small $\Delta T$ the inversion 
is produced by the de Jong mechanism, while without it 
the minimum $\Delta T$ to invert the population is about $\sim 1$ K. 
For small  $\fwd$, however, the inversion appears only 
for $\Delta T>15$ K. At higher $\Nh\sim 10^{10}\cminv3$ this limit 
is 3.5 K (see dashed curve in Fig.~\ref{fig:mastem}) which is very
close to the predictions of our four-level model (\ref{eq:deltmin}).

Fig.~\ref{fig:s2} shows the dependence of the optical depth in the main masing transition
$\taum$ on the  gas temperature for the same six sets of parameters $(\Nh,\fwd)$ as in
Fig.~\ref{fig:tauH}. One sees that in most points the
inversion appears when gas temperature exceeds $\Td$.  
 At point 4, however, the inversion exists even at smaller
$T$. This results from the action of the de Jong mechanism 
in an optically thin (in line) medium for a small dust content
(see also Fig.~\ref{fig:TH}). 

The optical depth increases sharply at small $\Delta T$ and
saturates at $T\sim500$ K. This behavior is accurately 
described by our analytical formula (\ref{eq:genii}). 
The optical depth is thus proportional to 
\be \label{eq:tautemp}
\taum\propto \left( {\rm e}^{-E_{41}/T} -
{\rm e}^{-E_{31}/\Td} \right) /\sqrt{T}, 
\ee
where the $\sqrt{T}$ factor comes from the Doppler width.

\subsection{Other maser transitions}

\begin{table}
\begin{center}
{\sc TABLE 1\\ Ortho-water \\ 
masing transitions}
\vskip 2pt
\begin{tabular}{rrr}
\hline
\hline
&\\
{\#} & {Transition} &{$\lambda\ \mum$} \\
 &\\
\hline
1 & $3_{12} \rightarrow 2_{21}$ & 260   \\ 
2 & $4_{14} \rightarrow 3_{21}$ & 789 \\
3 & $4_{23} \rightarrow 3_{30}$ & 669  \\ 
4 & $6_{16} \rightarrow 5_{23}$ & 13475 \\
5 & $5_{32} \rightarrow 4_{41}$ & 483   \\
6 & $6_{25} \rightarrow 5_{32}$ & 227  \\
7 & $6_{34} \rightarrow 5_{41}$ & 259   \\
8 & $6_{43} \rightarrow 5_{50}$ & 683   \\
9 & $8_{27} \rightarrow 7_{34}$ & 231   \\
10 & $7_{43} \rightarrow 6_{52}$ & 235   \\
11 & $7_{52} \rightarrow 6_{61}$ & 677   \\  
12 & $10_{29}\rightarrow 9_{36}$ & 933   \\
\hline
\end{tabular}
\label{tab:maser}
\end{center}
\end{table}

\begin{table}
\begin{center}
{\sc TABLE 2\\ Para-water \\ 
masing transitions}
\vskip 2pt
\begin{tabular}{rrr}
\hline
\hline
&\\
{\#} & {Transition} &{$\lambda\ \mum$} \\
 &\\
\hline
1 & $3_{13} \rightarrow 2_{02}$ & 1636  \\ 
2 & $5_{15} \rightarrow 4_{22}$ & 922 \\
3 & $5_{24} \rightarrow 4_{31}$ & 309  \\ 
4 & $5_{33} \rightarrow 6_{06}$ & 632 \\
5 & $6_{33} \rightarrow 5_{42}$ & 194  \\
6 & $7_{26} \rightarrow 6_{33}$ & 208  \\
7 & $6_{42} \rightarrow 5_{51}$ & 637   \\
8 & $7_{35} \rightarrow 6_{42}$ & 170  \\
9 & $7_{44} \rightarrow 6_{51}$ & 256  \\
10 & $9_{28} \rightarrow 8_{35}$ & 331   \\
11 & $8_{44} \rightarrow 7_{53}$ & 139   \\  
12 & $8_{53}\rightarrow 7_{62}$ & 252   \\
\hline
\end{tabular}
\label{tab:para}
\end{center}
\end{table}

Our simulations show that in addition to the $6_{16} \rightarrow 5_{23}$ maser, 
there appear many other masers.
The strongest masers for ortho-water are listed in Table~1 and  for para-water
in Table~2. 
The levels of the constant optical depth are 
shown in Figs.~\ref{fig:allmas} and \ref{fig:allpara}, respectively.
Some of these masers were found in the calculations of
\citet{CK84a} and \citet{WW97}. 
Most of them appear also in models based on the  \citet{dJ73}
mechanism \citep[see also][]{CE85,NM91}.
The $4_{14} \rightarrow 3_{21}$ and $3_{13} \rightarrow 2_{20}$ transitions have 
been detected in the Orion cloud by \citet{PK80} and \citet{Wa80}.
The masing emission in $10_{29} \rightarrow 9_{36}$ line was observed 
in a wide variety  of sources where 1.35 cm maser was detected
\citep{MMP90}.

\begin{figure}[htb]
\centerline{\epsfig{file=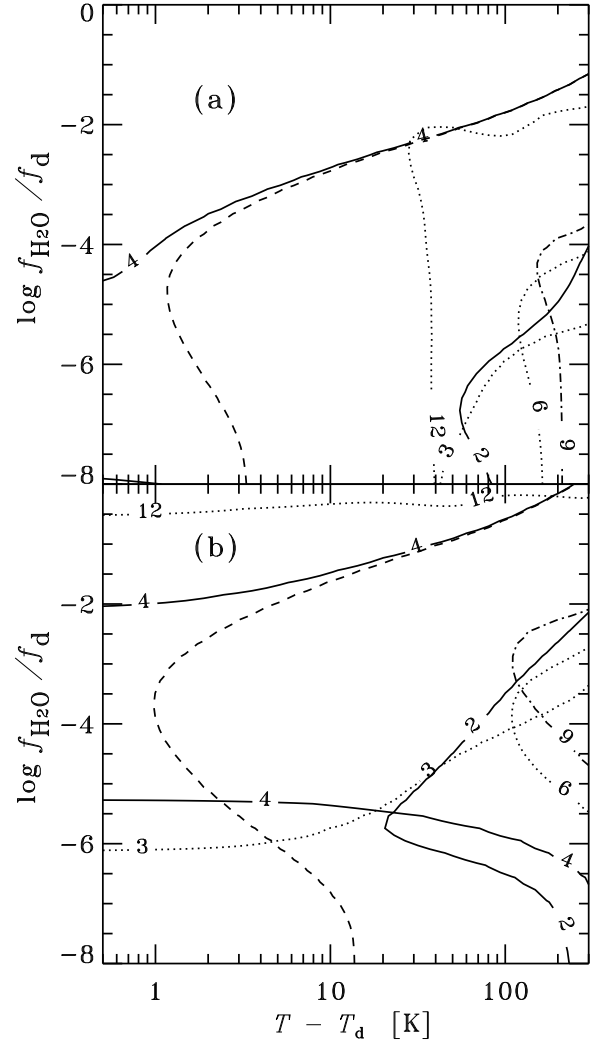,width=7.5cm}}
\caption{
Isolines of the zero inversion of the strongest masing  transitions
for the same parameters as in Fig.~\ref{fig:s3}. 
The labels on the contours correspond to the numbering in Table~1.
Panel
(a) is for $\Nh=10^{10}\cminv3$; (b) corresponds to $\Nh=10^9\cminv3$.
The inversion of the maser level populations exists
on the right side of the corresponding isoline.
The left dashed curves show the zero inversion for the $6_{16}\rightarrow 5_{23}$
maser when the de Jong mechanism is turned off.
}
\label{fig:mastem}
\end{figure}

\begin{figure}[htb]
\centerline{\epsfig{file=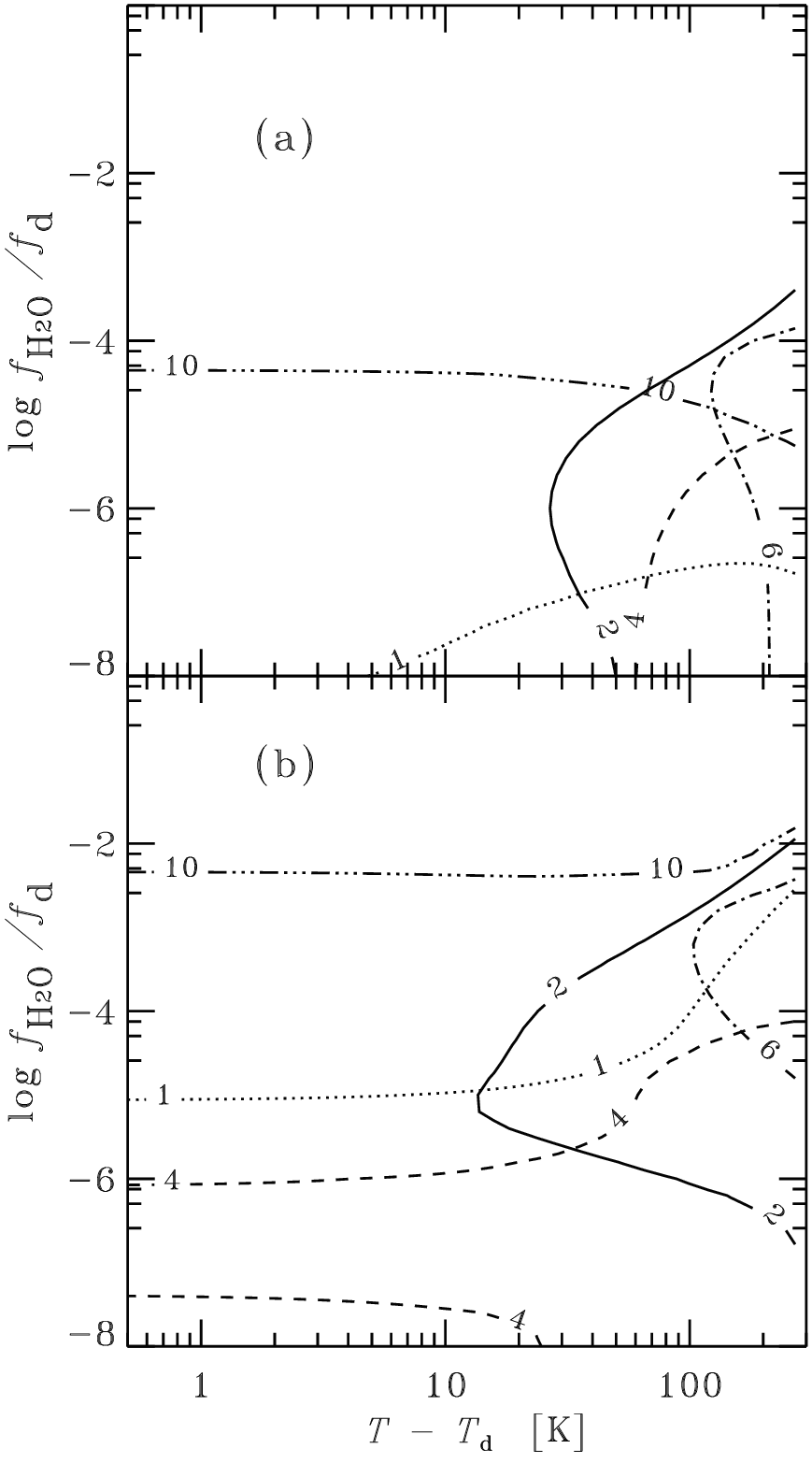,width=7.5cm}}
\caption{
Same as Fig.~\ref{fig:mastem}, but for para-water.
The labels on the contours correspond to the numbering in Table~2.
}
\label{fig:paratem}
\end{figure}

We show the   isolines of the zero inversion of the level populations
as a function of the gas temperature for the fixed $\Nh$
in Figs.~\ref{fig:mastem} and \ref{fig:paratem}.
The inversion of the level populations appears on the right
side from the corresponding curves.
We see that the inversion in most of the transitions  arises when the difference
between the gas and dust temperature is larger than 30--100 K and
one does not expect to see a significant signal in other 
water masing transitions if $\Delta T$ is small.
This is  related to the fact that the energy separation between the
corresponding levels is much larger than that 
for the $6_{16} \rightarrow 5_{23}$ maser.
Many masers also require the water concentration  two orders of magnitude smaller than
that where the $6_{16} \rightarrow 5_{23}$ and $10_{29} \rightarrow 9_{36}$
masers are strongest (see Figs.~\ref{fig:allmas} and \ref{fig:allpara}).
In these respects the cold dust -- hot gas model significantly differs
from the de Jong model where many masing transitions appear simultaneously.

\subsection{Other dust types and sizes}

Since the optical properties of the ice in the far infrared spectral range
for different grain size $0.01\ \mum \leq a \leq 1\ \mum$ are very similar,
the maser strength is almost independent of $a$.
The crystalline  ice has a stronger peak in $\kabs$ at $45\ \mum$ and
therefore the $6_{16}-5_{23}$ maser optical depth is 30\% larger
than that for the amorphous ice.
We repeated calculations with $0.1\ \mum$ silicate and graphite grains.
The optical depth decreases by 15\% and 25\%, respectively.

We also computed the maser conditions
for the mixture of the graphite and SiC dust.
We consider the size distribution of the dust grains
$\d n(a) \propto a^{-1.5}\d a$,
which extends from $a_{\min}=0.005\ \mum$ to $a_{\max}=10\ \mum$.
To allow easy comparison with the previous results  \citep{CW95,WW97},
we use an approximation of the results by \citet[][see their Fig. 6]{LD93} 
for the  dust cross-section.
We assume the dust opacity  in the form
$\kabs =600 (\fd/0.01)$ for $\lambda<50\ \mum$
and $\kabs =600 (\fd/0.01)(50\ \mum/\lambda)^2\ [\cmtwo\ginv]$, for $\lambda>50\ \mum$.
The resulting maser optical depth is very similar to that for the crystalline  ice.

\subsection{Approximating the inversion efficiency for $6_{16}\rightarrow5_{23}$
maser}

The maser luminosity is a function of the optical depth $\taum$ which
depends on the inversion $\delnm$. In many astrophysical problems, 
one would like to make simple estimations of 
the inversion and optical depth not repeating cumbersome calculations
of the water molecular level populations. Thus one would like to have
simple analytical approximate formulae for this purpose.

It is possible to design a formula for the inversion
that describes it with an error of a factor of two using the four-level
model from Sect.~\ref{sec:anal}. We can further simplify the model
by assuming that the upper masing level $6_{16}$ is populated
directly from the $4_{14}$ level.
We then arrive to a system of two equations, each corresponding to the
two-level model.
We now can introduce a pseudo-transition $1\rightarrow4$, and
prescribe to it some collisional and radiative rates.

We propose the following formula to compute the inversion:
\be \label{eq:appinv}
\delnm = \frac{n_1}{g_1} \left( \frac{S_{41}}{1+S_{41}} -
\frac{S_{31}}{1+S_{31}} \right) .
\ee
The population at the lower level 1 is given by the following
expression
\be
\frac{n_1}{g_1} = 0.75 U^{-1}_{\rm part} \exp(-289/T),
\ee
where,  based on the results of simulations,
we can approximate  the partition function in the range $10<T<500$ K as
\be
U_{\rm part}=2.86[1+(T/42)^{1.4}] .
\ee
Approximating the dependence of the collisional rate on the gas temperature
by a power-law, we get the ratio
\be
\epsilon_{31}=0.033 \frac{\Nh}{10^{10}} \left(\frac{T}{100}\right)^{0.92}
\left( 1-\exp[-319/T] \right) .
\ee
The dust influence is parametrized by $\beta$:
\be
\beta_{31}=2\times 10^{-6} (T/100)^{1/2} (\fwd)^{-1} ,
\ee
and $\delta_{31}$ is computed using equation (\ref{eq:delapp}).
We now can assume that the $4\rightarrow1$ transition is characterized by
$\epsilon_{41}=1.1 \epsilon_{31}$ and  $\beta_{41}=0.6 \beta_{31}$.
The source functions are found from equation (\ref{eq:sbb}).
These formulae give a rather good  agreement with the results
of numerical simulations (within a factor of two).

\section{Conclusions}
\label{sec:concl}

We consider the cold dust -- hot gas mechanism of pumping of the water masers. 
To obtain the inversion of the maser level populations 
we take into account the first 45 rotational levels of ortho- and para-\water\ 
molecules and all possible collisional and radiative transitions between them.
In the case of the large optical depth in the main pumping lines
and the unsaturated masers,
the radiative transfer problem can be significantly simplified, allowing us to
investigate the maser efficiency for the set of the main parameters,
such as the hydrogen concentration, the gas-to-dust mass ratio and
the gas temperature. 

As it was suggested by \citet{De81}, the pumping cycle  $4_{14} \rightarrow 5_{05}
\rightarrow 6_{16} \rightarrow 5_{23} \rightarrow 4_{14}$
mainly determines the inversion of 
the $6_{16}-5_{23}$ maser levels populations.
We use this four-level model for the analysis of our numerical results. 
We also suggest approximate formulae for the dependence of the inversion
on the hydrogen concentration, the gas-to-dust mass ratio and
the gas and dust temperatures, that  could be used for
modeling  the astrophysical sources.
We find that the inversion is the largest
when the water-to-dust mass ratio is $\fwd\sim10^{5}/\Nh$, and the 
maser optical depth is the largest when $\Nw\sim(10^{6}\div 10^{7})\fd$.

The inversion of the maser level populations as a function of the same parameters is
calculated in the case of the amorphous and crystalline ice, silicate as well as
graphite dust. The results of these calculations are very similar.
Because the inversion depends on the dust mass density, our results
can be applied to any size distribution of the dust particles.

Our analysis shows that in the optically thick environments 
the inversion in the $6_{16} \rightarrow 5_{23}$
transition appears when the gas is just $\sim 1$ K hotter than the dust,
while most of other transitions start to be
inverted at much larger $\Delta T\sim 30-100$ K. Most of them require also
smaller water-to-dust mass ratio $\fwd$. 
This is very different from the predictions
of the de Jong model where many transitions are inverted simultaneously.
If the cold dust -- hot gas model is the correct one,
the relative ratios of the luminosities in different
masing transitions could provide constraints on the
gas and dust temperatures,  water-to-dust mass ratio, and hydrogen concentration.

\begin{acknowledgements}
This work was supported by the Centre for International Mobility,
the Magnus Ehrnrooth Foundation, the Finnish Graduate School for Astronomy and
Space Physics, and the Academy of Finland.
We are grateful to Ryszard Szczerba for providing the dust absorption
coefficients, to Dmitrii Nagirner for providing the codes for computations of
$K$- and $L$-functions, and to Vsevolod Ivanov for useful comments on the
manuscript and discussions.
\end{acknowledgements}


\end{document}